\documentclass[preprint2]{aastex}
\usepackage[hyperindex,breaklinks]{hyperref}

\newcommand{\cxo}{{\it Chandra}}

\newcommand{\hst}{{\it HST}}

\newcommand{\galex}{{\it GALEX}}
\newcommand{\spitzer}{{\it Spitzer}}
\newcommand{\etal}{et al.}

\newcommand{\nh}{\mbox {$N_{\rm H}$}}

\newcommand{\ha}{H{\sc $\alpha$}}

\newcommand{\hi}{H\,{\sc i}}
\newcommand{\hii}{H\,{\sc ii}}
\newcommand{\sii}{[S\,{\sc ii}]}
\newcommand{\nii}{[N\,{\sc ii}]}

\newcommand{\oiii}{[O\,{\sc iii}]}

\newcommand{\about}{$\sim$\kern.03em}

\newcommand{\chase}{ChASeM33}
	\newcommand{\m}{M33}

\slugcomment{Draft \today}

\shorttitle{\chase}
\shortauthors{T\"ullmann et al.}

\begin{document}


\title{The \cxo\ ACIS Survey of \m\ (\chase):\\
Investigating the Hot Ionized Medium in NGC604}


\author{Ralph T\"ullmann\altaffilmark{1},
Terrance J. Gaetz\altaffilmark{1},
Paul P. Plucinsky\altaffilmark{1},
Knox S. Long\altaffilmark{2},
John P. Hughes\altaffilmark{3},
William P. Blair\altaffilmark{4},
P. Frank Winkler\altaffilmark{5},
Thomas G. Pannuti\altaffilmark{6},
Dieter Breitschwerdt\altaffilmark{7},
and
Parviz Ghavamian\altaffilmark{4}}
\altaffiltext{1}{Harvard-Smithsonian Center for Astrophysics, 60 Garden Street, Cambridge, MA 02138; rtuellmann@cfa.harvard.edu}
\altaffiltext{2}{Space Telescope Science Institute, 3700 San Martin Drive, Baltimore, MD 21218}
\altaffiltext{3}{Dep. of Physics and Astronomy, Rutgers University, 136 Frelinghuysen Road, Piscataway, NJ 08854}
\altaffiltext{4}{Dep. of Physics and Astronomy, Johns Hopkins University, 3400 North Charles Street, Baltimore, MD 21218}
\altaffiltext{5}{Dep. of Physics, Middlebury College, Middlebury, VT 05753}
\altaffiltext{6}{Space Science Center, Morehead State University, 200A Chandler Place, Morehead, KY 40351}
\altaffiltext{7}{Institut f\"ur Astronomie, Universit\"at Wien, T\"urkenschanzstr. 17, A-1180 Wien, Austria}


\begin{abstract}
NGC604 is the largest \hii-region in M33, second only within the Local Group to 30\,Dor, and is important as a laboratory for understanding how massive young stellar clusters interact with the surrounding interstellar medium. Here, we present deep (300ks) X-ray imagery of NGC604 obtained as part of the \cxo\ ACIS Survey of M33 (ChASeM33), which show highly structured X-ray emission covering $\sim$70\% of the full \ha\ extent of NGC604. The main bubbles and cavities in NGC604 are filled with hot ($kT=0.5$\,keV) X-ray emitting gas and X-ray spectra extracted from these regions indicate that the gas is thermal. 
For the western part of NGC604 we derive an X-ray gas mass of $\sim$4300\,$M_{\odot}$ and an unabsorbed (0.35\,--\,2.5\,keV) X-ray luminosity of $L_{\rm X}=9.3\times 10^{35}$\,erg\,s$^{-1}$. These values are consistent with a stellar mass loss bubble entirely powered by about 200 OB-stars. This result is remarkable because the standard bubble model tends to underpredict the luminosity of X-ray bright bubbles and usually requires additional heating from SNRs. Given a cluster age of $\sim$\,3\,Myr it is likely that the massive stars have not yet evolved into SNe.
We detect two discrete spots of enhanced and harder X-ray emission, which we consider to be fingerprints from a reverse shock produced by a supersonic wind after it collided with the shell wall. 
In the eastern part of NGC604 the X-ray gas mass amounts to $\sim$1750\,$M_{\odot}$. However, mass loss from young stars cannot account for the unabsorbed X-ray luminosity of $L_{\rm X}=4.8\times 10^{35}$\,erg\,s$^{-1}$. Off-center SNRs could produce the additional luminosity. The bubbles in the east seem to be much older and were most likely formed and powered by young stars and SNe in the past. A similar dichotomy between east and west is seen in the optical, implying that a massive wall of neutral and ionized gas shields the dynamically quiescent east from the actively star forming west.
\end{abstract}




\section{Introduction}
NGC604 is the second largest giant \hii-region in the Local Group and has been studied throughout all wavelengths, from radio \citep{church99,tosa07}, infrared \citep{hig03}, optical \citep{rosa82, teno00}, ultraviolet \citep{rosa80,keel04}, to the X-ray regime \citep{apel04}. The main focus of these studies was the investigation of the effects that massive young stars impose on the surrounding multi-phase interstellar medium.

It is commonly believed that NGC604 is ionized by the radiation field and winds from a young and very massive stellar population. A detailed analysis of the stellar content indicates the existence of about 200 O and WR-stars with a WR/O-ratio of about 0.075 \citep{dris93,hunt96,gonzo00,bruh03}. Although the stellar content, the gas mass, and the average gas density of NGC604 are different compared to other giant \hii-regions in, e.g., the LMC, II\,Zw\,40, or NGC4214 \citep{teno06}, the blister-like structures and overall \ha\ morphology resulting from stellar winds and/or SNe activity is very similar among all these regions. In this regard the similarity with 30\,Dor in the LMC is the most striking \citep[see, e.g.,][]{wang99,town06}. 

Besides the warm ionized phase of the ISM in NGC604, the hot ionized medium (HIM) is another crucial, yet poorly investigated, phase of the ISM which tells us much about the internal energetics, the wind-ISM interaction, the evolution of the bubble, and the sources which power it. 
A difficulty in the past originated from the fact that several X-ray bright superbubbles ($\log L_{\rm X}>35$) in the LMC show an excess X-ray luminosity in comparison to predictions from the standard bubble model \citep{cas75,weaver77}. There is general agreement that this excess emission can be attributed to off-center SNRs interacting with the shell walls \citep{chu90,wang91}. On the other hand, X-ray dim bubbles are found to be consistent with the standard model and do not require additional heating from SNRs \citep{chu95}. Another subdivision of stellar bubbles was introduced by \citet{oey96} who distinguished between high velocity bubbles, i.e., those whose expansion velocity is underestimated by the standard model and those which are consistent with it. The emerging picture seems to be that X-ray bright, high velocity bubbles are additionally heated by SNRs and that X-ray dim, slowly expanding bubbles do not require additional heating.

With the current study, which is part of the \cxo\ ACIS Survey of M33 \citep[ChASeM33,][]{plu08}, we present the first detailed X-ray analysis of NGC604 in order to explore the origin of the X-ray emission and to test the wind-blown bubble scenario. Although NGC604 \citep[$D$\,=\,817\,kpc,][]{freed01} has been detected in previous major X-ray surveys \citep{long96,misa06,plu08}, the only significant observation that has been published is a \cxo\ contour map of the central part \citep{apel04},  which revealed that the bubbles seen in the optical are filled with X-ray emitting gas. However, the limited sensitivity from this 90\,ks observation (ObsID \dataset[ADS/Sa.CXO#obs/02023]{2023}) inhibited further analysis. 

The paper is structured as follows: after an outline of the data reduction steps for imaging (Sect. 2.1) and spectroscopy (Sect. 2.2), we present the main results derived from both techniques in Sect. 3. In Sect. 4.1 we discuss the origin of the X-ray gas. By means of spectral modeling, we determine electron densities and filling factors of the HIM and derive X-ray gas masses for individual regions (Sect. 4.1.1). In Sects 4.1.2 and 4.1.3, we cross-check whether the stellar O/WR population in NGC604 can account for the observed X-ray gas masses via continuous stellar mass loss and if the observed X-ray luminosity is consistent with the one predicted from the ``standard'' model of a stellar wind-blown bubble \citep{weaver77}. Sect. 4 is concluded by evaluating whether SNe produced a significant fraction of the observed X-ray gas mass (Sect. 4.2) and by presenting a consistent picture of NGC604 which emerges from the multi-wavelength data analysis (Sect. 4.3). A summary is provided in Sect. 5.

\section{Observations and Data Reduction}
This study is part of the \cxo\ ACIS Survey of M33 (ChASeM33) and presents the first deep and high resolution X-ray images of the diffuse emission of \object{NGC604}. We chose 2 different ACIS-I fields, Field\,2 from ChASeM33 (using ObsIDs \dataset[ADS/Sa.CXO#obs/06378]{6378}, \dataset[ADS/Sa.CXO#obs/06379]{6379}, and \dataset[ADS/Sa.CXO#obs/07402]{7402}, \about 4\arcmin\ off-axis) and one archival field (ObsID 2023, with NGC604 at the aim point; see also dashed circle in Fig.~1 of \citet{plu08}). The total integration time was 300\,ks. 

All \chase\ observations were performed with ACIS-I in `Very Faint' (VFAINT) mode, whereas the archival one was obtained with ACIS-I in 'FAINT' mode. All data were reprocessed with CIAO version 3.4.1.1 and CALDB version 3.3.0.1.
 
The basic data reduction was carried out by following standard procedures. First, a new {\tt level=2} eventfile was created with {\tt acis\_process\_events} by applying new background flags for the {\tt VFAINT} and {\tt FAINT} modes, correcting for charge transfer inefficiency (CTI) and time-dependent gain variations. Pixel randomization was removed to avoid degrading the resolution of the data. The resulting event lists were filtered for events with {\tt grade=0,2,3,4,6} and {\tt status=0}, and the pipeline-generated good time intervals (GTI) were applied. No aspect solution correction was needed\footnote{see http://cxc.harvard.edu/ciao3.4/threads/arcsec\_correction/}. Positional uncertainties are $\le 1\arcsec$.

Subsequently, we checked for background flares by extracting a background light curve from the events on the ACIS-I3 chip, after removing point sources detected by the task {\tt wavdetect}. An iterative sigma-clipping algorithm was used to remove time intervals with count rates more than $3\sigma$ from the mean of each iteration, until all count rates were within $\pm3\sigma$ of the mean.\footnote{http://cxc.harvard.edu/ciao3.4/threads/filter\_ltcrv/} The GTI-corrected total exposure time was 294\,ks. 

\subsection{Imaging}
We created X-ray images for each ObsID in the following diagnostic energy bands:  0.35\,--\,8.0keV (broad), 0.35\,--\,1.1\,keV (soft), 1.1\,--\,2.6\,keV (medium), and 2.6\,--\,8.0~keV (hard).
Individual exposure maps for each ObsID and energy band were computed using the CIAO tools {\tt mkinstmap} and {\tt mkexpmap}. 
In order to ensure that the number of counts in each pixel of the exposure corrected image is proportional to the integrated number of counts of the incident source spectrum (over a given energy band), 
weighted instrument maps\footnote{http://cxc.harvard.edu/ciao3.4/threads/spectral\_weights/} were created assuming a spectrum representative of the diffuse X-ray emitting gas. We assumed an {\tt APEC} thermal plasma model with $kT=0.4$\,keV \citep{long96} and an averaged column density of \nh\ = $1.1\times10^{21}$~cm$^{-2}$ \citep{church99,lebo06}. 
The script {\tt merge\_all} was used to produce a merged event list and merged exposure maps for the different energy bands. Merged images with integer counts (Fig.~\ref{f1}) were created with {\tt dmcopy} after applying {\tt bin} and {\tt energy} filters for the individual energy bands.
Exposure-corrected images were created by dividing the (merged) images by their associated (merged) exposure maps. 
All images were binned by 4 pixels to a resolution of \about2\arcsec\ to gain sensitivity and to compensate for varying PSFs at different off-axis angles. The X-ray images which were used for the contour maps (Fig.~\ref{f2}) and the RGB-image (Fig.~\ref{f3}) were additionally smoothed in {\tt DS9} \citep{joy03} with a Gaussian kernel radius of 2 pixels, to help enhance the contrast between background and diffuse emission. 

For a multi-wavelength comparison of the different phases of the ISM in NGC604, we also employed archival data collected at near infrared (NIR, \spitzer/IRAC, 8$\mu$m), optical (\hst/WFPC2, \ha, \nii), and far ultraviolet (FUV, \galex, 1500\AA) wavelengths. Because we use the supplemental images exclusively for morphological comparison, standard data pipeline calibration was applied. The \hst\ \ha\ image was also continuum subtracted using a scaled version of the V-band image.

\subsection{Spectroscopy}
Because we are mainly interested in the purely diffuse emission of NGC604, we removed point sources from every event list using the {\tt ACIS Extract} \citep{Broos02} regions from the \chase\ point source catalog \citep{plu08}. For proper background removal, we ensured that the background regions (boxes) were located at the same distance from the readout node as our source regions (to account for the row dependence of front-illuminated CCDs). In order to improve photon statistics for the background data, we chose the sizes of these background regions to be approximately three times the size of the source region, yielding typically \about1200\,cts. 
 
The NGC604 pointings (Field\,2 and the archival one) differ in pointing direction and roll. We therefore did not produce a single merged event list and instead extracted source and background spectra from each individual event list. For this purpose we used {\tt specextract}, which also creates custom response matrix files (RMFs) and appropriate auxiliary response files (ARFs). The merging of RMFs and ARFs was done using the {\tt FTOOLS} tasks {\tt addarf} and {\tt addrmf} and by applying a weighting factor based on the GTI-corrected exposure time taken from the {\tt EXPTIME} header keyword. Finally, the merging of the spectra was accomplished with the task {\tt addspec}. We ensured that each spectrum contained the proper {\tt BACKSCAL}-keyword and was linked to its corresponding background-file. The source spectra were individually grouped with {\tt dmgroup} until a signal to noise ratio (S/N) of at least 3 per channel was achieved. All spectra were fitted with {\tt XSPEC} \citep{Arn96} {\tt ver.\,12.3.1}, using an {\tt APEC} collisional ionization equilibrium (CIE) model \citep{smith01}.

\section{Results}
\subsection{Imaging Analysis}
\subsubsection{Counts Images}
\begin{figure*}
\centering
\includegraphics[width=0.9\textwidth,angle=0]{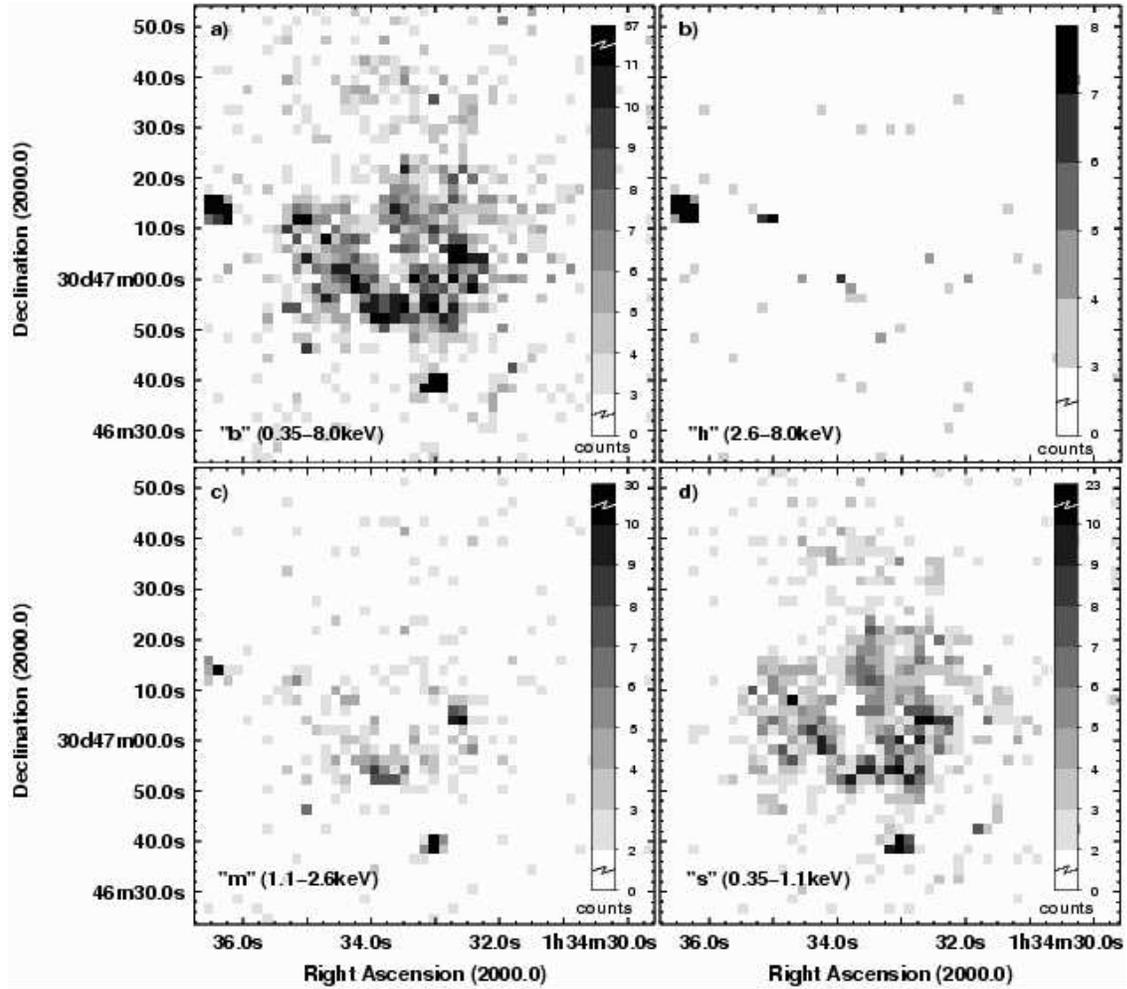}
\caption{\label{f1}
Counts images of NGC604 (binned by 4x4 pixels) for the broad (b), hard (h), medium (m), and soft (s) energy band shown on a linear gray scale. The integer counts in each band are plotted for the minimum and maximum count values neglecting the AGN source visible eastward of NGC604 in panels a\,--\,c. White represents pixel ranges belonging to the background (95\% Poission confidence). In case of NGC604 10\arcsec\ corresponds to 36\,pc.}
\end{figure*}

In Fig.~\ref{f1}, we show counts images of NGC604 for the different energy bands on a linear min/max-scale. These images are not exposure corrected and therefore preserve integer counts to allow an evaluation of the number of counts in each pixel. Pixels with observed counts consistent with background alone are shown in white (95\% Poission confidence). That is, assuming Poission statistics, we are 95\% confident that pixels with $\le2$ counts (panels a and b) and $\le 1$ counts (panels c and d) belong to the background.
\begin{figure*}
\centering
\includegraphics[width=0.9\textwidth,angle=0]{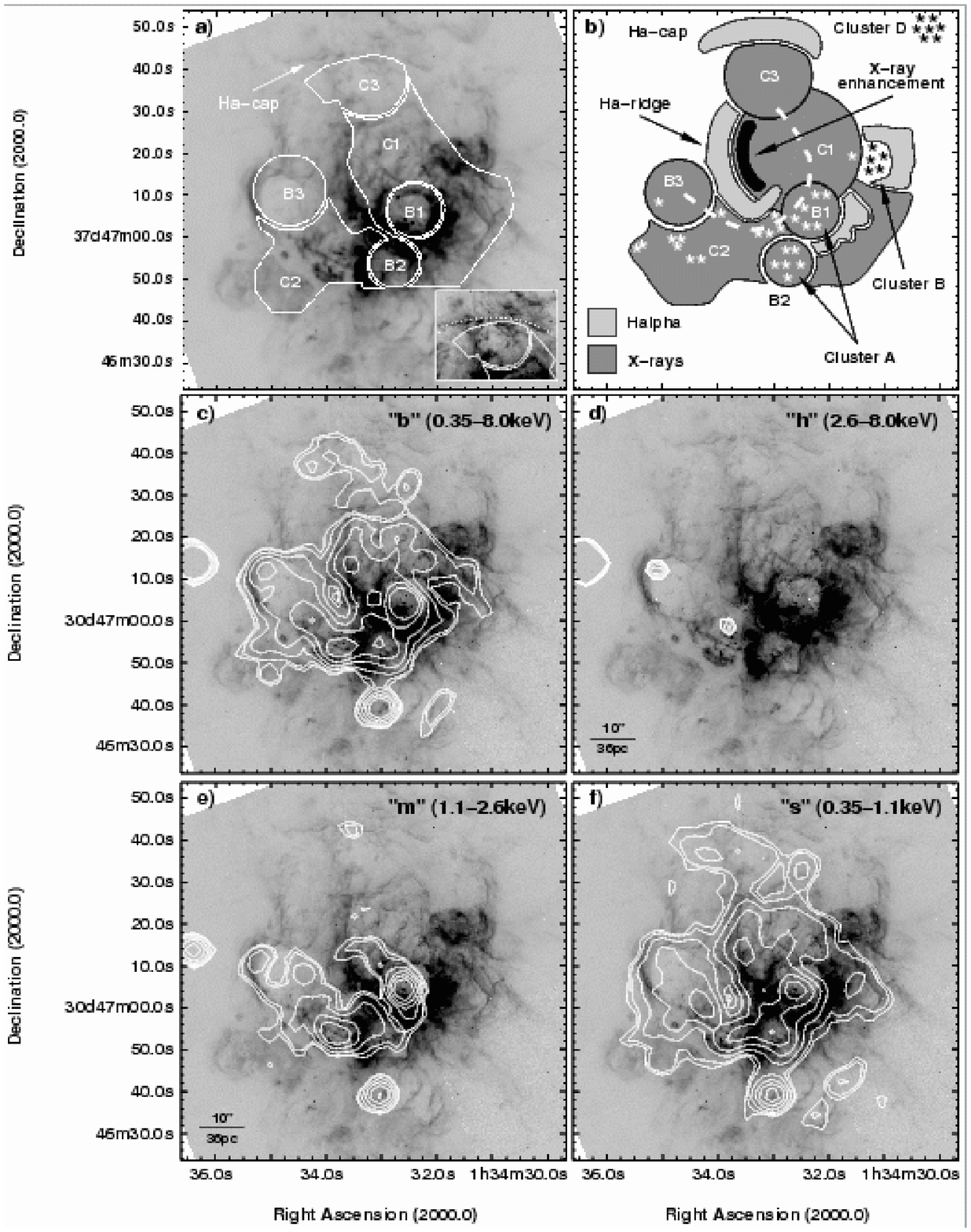}
\caption{\label{f2}
Panel a): Solid lines encircle regions with X-ray emission from which spectra were extracted. The dotted line in the inlay marks the location of the \ha-'cap'. Panel b): Sketch of NGC604, with regions and cavities defined in the text. The locations of stars and stellar clusters \citep{hunt96} and the 'U'-shaped morphology of the X-ray gas (white dashed line) are also indicated. Panels c)\,--\,f): Contour maps for the different energy bands (see text). In all panels, except panel b, the underlying image is the continuum-subtracted \hst\ \ha\ image.}
\end{figure*}

From these images it is obvious that NGC604 emits significant diffuse X-ray emission, most of it in the soft energy band, roughly forming the shape of an 'U'. There is also significant soft emission in the north at RA\,=\,01$^{h}$34$^{m}$$33.30^{s}$, Dec\,=\,+30\degr47\arcmin42\farcs0 (all coordinates in this paper are given for J2000.0), which was previously not detected by \citet{apel04}. Emission seen in the medium X-ray band largely coincides with that seen in the soft band, but does not reach the same extent. In the hard X-ray band, we detect only two significant emission features, one at RA\,=\,01$^{h}$34$^{m}$$35.00^{s}$, Dec\,=\,+30\degr47\arcmin12\farcs0 (15\, net cts) and the other at RA\,=\,01$^{h}$34$^{m}$$34.00^{s}$, Dec\,=\,+30\degr47\arcmin00\farcs0 (6\,net cts). The bright extended source visible in the broad band at RA = 01$^{h}$34$^{m}$36.50$^{s}$ and Dec = +30\degr47\arcmin15\farcs0 (\chase\ source 346) is most likely a background AGN \citep[see][for details]{plu08}. The southernmost source located at RA\,=\,01$^{h}$34$^{m}$33$^{s}$ and Dec\,=30\degr46\arcmin40\farcs0 can be identified with SNR GKL98-94 \citep{dodo80,gord98,gha05} and is listed as source number 342 in \citet{plu08}.
Another feature is the region of reduced X-ray emission running roughly north\,--\,south through the center of NGC604 and can best be seen in panels a and d.

\subsubsection{Contour Maps}
In Figs.~\ref{f2}c\,--\,\ref{f2}f, we provide X-ray contour maps for NGC604 based on exposure corrected images in the same energy bands as shown in Fig.~\ref{f1}. These maps were smoothed in {\tt DS9} using a kernel radius of 2 pixels to enhance the contrast between source and background and are overlaid on the \hst\ \ha\ image to allow a direct comparison with structures of the warm ionized gas. For each energy band, six contour levels are plotted on a sqrt-scale, starting at 3$\sigma$ above the background. The b-band contours cover $4.55\times10^{-8}$ to $1.67\times10^{-7}$\,counts/s/cm$^{2}$, the h-band covers $1.68\times10^{-8}$\,--\,$2.72\times10^{-8}$\,counts/s/cm$^{2}$, the m-band covers $6.69\times10^{-9}$\,--\,$3.3\times10^{-8}$\,counts/s/cm$^{2}$, and the s-band covers $2.41\times10^{-8}$ to $1.67\times10^{-7}$\,counts/s/cm$^{2}$.

The broad-band emission extends $\sim$\,250\,pc along the N-S axis and $\sim$\,240\,pc along the E-W axis, respectively. As can be seen from the \hst\ \ha\ image, the two hard X-ray sources as well as the X-ray-bright spot at medium energies do not have \ha-emitting counterparts. The lack of X-rays at the center of NGC604 is likely due to internal absorption (see below). SNR GKL98-94 is not detected at energies above 2.6\,keV. 
However, to identify all those features and to constrain the origin of the gas, it is necessary to discuss the \cxo\ observations in the framework of other multi-wavelength data. 
 
\subsubsection{Multi-wavelength Comparison}
\begin{figure*}
\centering
\includegraphics[width=0.9\textwidth,angle=0]{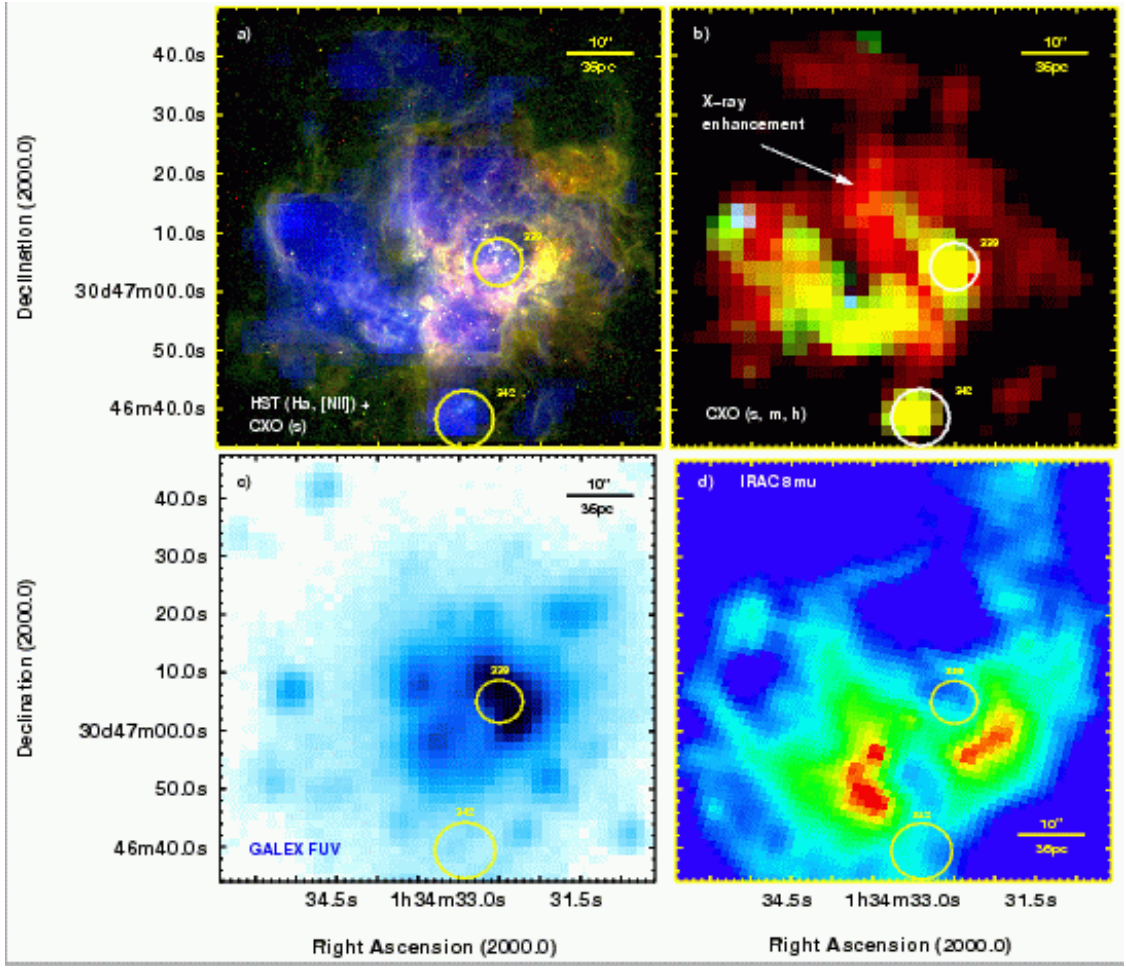}
\caption{\label{f3}
Panel a): A multi-wavelength comparison between \ha\ (red), \nii\ (green), and soft X-rays (0.35\,-1.1\,keV, blue) reveals that the bubbles visible in \ha\ are filled with hot gas and host a massive stellar population. It also shows the negative morphological correlation between the different gaseous phases. Panel b): This three color high resolution (2\arcsec) composite image provides the first detailed look of the X-ray emission and its spatial extent. Red represents the soft, green the medium, and blue the hard band, respectively. Panel c): The stellar continuum emission detected in the FUV indicates the existence of young stars and further strengthens the link between hot gas and stars as ionizing sources. Panel d): Three of the \ha-bright bubbles can also be seen in the NIR at $8\mu$m. The emission seen in this band traces molecular dust grains and implies that the interior of these structures has been excavated by stellar activity . Yellow circles are X-ray sources taken from the ChASeM33 source catalog \citep{plu08}.} 
\end{figure*}

The multicolor image in Fig.~\ref{f3}a combines narrowband \hst\ imaging (\ha\ and \nii) with the soft X-ray emission detected by \cxo. It unambiguously demonstrates that all bubbles and cavities are filled with hot X-ray emitting gas. The hot gas appears to bend around the \ha\ ridge \citep{teno00}, roughly forming a 'U'-shaped structure (cf. Fig.~\ref{f2}b). 
Fig.~\ref{f3} also shows the negative morphological and spatial correlation between soft X-rays and \ha-emitting gas. A similar negative correlation is also seen in comparing the soft X-ray emission with the 8.4\,GHz radio continuum map presented by \citet{church99}. This kind of morphological anticorrelation agrees well with our understanding of the assembly and composition of a multi-phase ISM. 
A much more detailed view of the complex morphology of NGC604 is provided by the first high-resolution (2\arcsec) three color X-ray image of NGC604 (Fig.~\ref{f3}b). On this image the 'U'-shaped soft X-ray emission is clearly visible, bending around the prominent ridge of ionized hydrogen and molecular gas (Fig.~\ref{f3}d). 

As pointed out by \citet{teno00}, the ionization structure (and thus the radiation field) as well as the velocity field in the western part is completely different from that in the eastern part. This can be attributed to the presence of the \ha\ ridge which may be shielding the east from the violent stellar evolutionary processes and the hard radiation field in the west. This kind of dichotomy is also reflected by the distribution of the FUV emission (Fig.~\ref{f3}c) which is a tracer of the stellar continuum emitted by hot stars. The bulk FUV emission originates westwards of the \ha\ ridge and coincides with the northern part of star cluster A \citep{hunt96} which contains about 200 O/WR stars. Considering the stellar winds of such stars, it appears to be reasonable to attribute the existence of the X-ray emission in that region to stellar mass loss. However, the X-ray emission in the north and east of NGC604 is unlikely to be explained by the same kind of ionization mechanism, simply because of the absence of similarly massive clusters.

As there is no significant hot gas detectable beyond the outer extent of the warm ionized gas, one might think of NGC604 as a giant boiling \hii-region whose bubbles partly broke out, merged, and formed the observed blister-like structures. The released X-ray gas, however, still has not yet penetrated the outer \hii-shell. 

\subsubsection{Individual Regions}
Some of the X-ray emitting regions found by our survey overlap with areas investigated in the optical by \citet{teno00}, \citet{apel04}, and \citet{hunt96}. Because these studies did not use a common nomenclature and the borders of the investigated areas are unknown in some cases, and because our study also incorporates new regions, we decided to define our own regions and nomenclature. 
The interesting areas are highlighted by solid white lines in Fig.~\ref{f2}a, and encircle the emission detected in the broad-contour map (Fig.~\ref{f2}c). Shapes and sizes, i.e. the assumed geometries, of the emitting regions are shown as white dashed lines and were estimated by examining the available multi-wavelength data. Obvious examples are bubbles B1\,--\,B3 whose \ha\ morphology implies a roughly spherically-symmetric geometry. We also investigate some non-spherical regions which we call cavities (C1\,--\,C3). These cavities are also important, because the bubbles seem to be either connected to or embedded in these regions, suggesting that the cavities are also influenced by the stellar wind and radiation field. A nice example is B1 which seems to be part of the much larger cavity C1 (see Fig.~\ref{f2}b for a sketch of NGC604). Cavities C2 and C3, finally, are of interest because they contain X-ray emission which apparently belongs to neither of the other regions.

Regions B1 and B2 appear to be coincident with regions 1 and 2 investigated by \citet{teno00}. The same areas are labeled 'A' and 'B' in \citet{apel04}. Moreover, region 'C' and parts of region 'D' in \citet{apel04} correspond to our regions B3 and C1, respectively. Most of the stars investigated by \citet[][their clusters A and B]{hunt96} are coincident with our regions B1, B2, and C1. Cluster A covers regions B1 and B2 while cluster B is coincident with region C1.
Regions C2 and C3 have no corresponding counterparts and were newly introduced by us.
In the following we briefly comment on the most important observational results for each of these regions.
\paragraph{Cavity C1:}
From Fig.~\ref{f3}a it appears that this region represents the largest coherent structure in NGC604 formed by stellar activity. C1 hosts bubble B1 which contains the bulk of cluster A. In addition, it also harbors cluster B which contains about 8 O9-stars (see Fig.~\ref{f2}b for its location). As the multi-color X-ray image of Fig.~\ref{f3} implies the whole cavity is filled with soft X-ray emitting gas. The noticeable depression at RA\,=\,01$^{h}$34$^{m}$$32.0^{s}$, Dec\,=\,+30\degr47\arcmin10\farcs0 is apparently caused by absorption along the B1 shell. 

The diffuse X-ray emission is significantly enhanced westward of the \ha\ ridge, nicely following that structure. This feature is also detected at energies above 1.1\,keV and indicates significantly higher plasma temperatures.
\paragraph{Cavity C2:}
Soft X-ray emission is also clearly detected in C2, a region in the eastern part of NGC604, which separates B1 and B2 from B3. No emission is visible on the X-ray map presented by \citet{apel04}. The western part of C2 contains a significant fraction of the stellar content of cluster A. This area also emits at medium X-rays, as can be seen in Fig.~\ref{f2}. From the multi-wavelength data we identify 3 stellar clusters in the eastern part of C2 (see Fig.~\ref{f2}b for their positions).
\paragraph{Cavity C3:}
Another striking feature in NGC604 visible in Fig.~\ref{f2}f is the soft diffuse X-ray emission found in the northernmost region C3 (cf. also Fig.~\ref{f3}b). This emission was previously not detected at a significant level by \citet{apel04} and appears to be confined towards the north by a 'cap'-like structure seen in \ha\ (see Figs.~\ref{f2}a and \ref{f2}b). The full extent of the X-ray and \ha\ emission can best be seen from Fig.~\ref{f2}c. 

Another feature detected at medium energies is located north of the soft X-ray source at RA = 01$^{h}$34$^{m}$$33.5^{s}$, Dec\,=\,+30\degr47\arcmin44\farcs0. From our multi-wavelength data no point sources are visible which could be responsible for the detected X-rays. It should be pointed out that the X-ray gas in C3 fills up the gap in the radio continuum distribution observed by \citet[see][]{church99}. 
\paragraph{Bubble B1:}
Bubble B1 is adjacent to the largest cavity C1 and contains a substantial fraction of cluster A. A multi-wavelength comparison between FUV, X-ray, optical, and NIR emission (Fig.~\ref{f3}) shows that \chase\ source No. 339 is coincident with the stars located in B1. These cluster stars are also a source of strong FUV emission and can be associated with extended thermal IR emission, which also hints at the presence of a young stellar population. In addition, this region also emits strongly in soft and medium X-rays.

\begin{table*}[pht]
\caption{\label{t1} Best fit parameters from the multi-temperature CIE model}
\begin{tabular}{ccccc|cc|ccc}
\hline\hline
ID  & \# of  & $\chi^{2}_{\rm red}$ &dof& nhp   & $K_{1}$         & $K_{2}$                   & $kT_{1}$               & $kT_{2}$               & $kT$              \\
    & cnts   &         &       &      &     \multicolumn{2}{c|}{$10^{-5}\,{\rm cm^{-5}}$}      &                 \multicolumn{3}{c}{keV}                             \\
(1) & (2)    & (3)     &   (4) & (5)  &     \multicolumn{2}{c|}{(6)}                           &                 \multicolumn{3}{c}{(7)}                             \\
\hline
C1  &  645   & 0.723   & 16    & 0.77 & 4.190$^{+0.469}_{-0.964}$ & 0.687$^{+0.257}_{-0.186}$ & 0.22$^{+0.04}_{-0.03}$ & 0.80$^{+0.17}_{-0.10}$ & 0.40$^{+0.06}_{-0.05}$
 \\
C2  &  452   & 0.801   & 18    & 0.70 & 1.573$^{+0.570}_{-0.378}$ & 0.580$^{+0.087}_{-0.122}$ & 0.27$^{+0.05}_{-0.04}$ & 1.01$^{+0.26}_{-0.04}$ & 0.55$^{+0.13}_{-0.04}$ \\
C3  &  162   & 0.247   & 10    & 0.99 & 0.432$^{+0.265}_{-0.178}$ & 0.206$^{+0.280}_{-0.094}$ & 0.27$^{+0.18}_{-0.25}$ & 0.69$^{+0.29}_{-0.17}$ & 0.44$^{+0.23}_{-0.21}$ \\
B1  &  109   & 0.479   & \ \,6 & 0.82 & 2.881$^{+0.558}_{-1.019}$ & 0.245$^{+0.103}_{-0.234}$ & 0.18$^{+0.53}_{-0.21}$ & 0.61$^{+0.11}_{-0.23}$ & 0.28$^{+0.39}_{-0.22}$ \\
B2  &  126   & 0.560   & \ \,7 & 0.79 & 1.167$^{+0.085}_{-0.796}$ & 0.111$^{+0.231}_{-0.104}$ & 0.22$^{+0.08}_{-0.06}$ & 0.70$^{+0.23}_{-0.13}$ & 0.33$^{+0.11}_{-0.08}$ \\
B3  &  229   & 0.437   & 15    & 0.97 & 0.845$^{+0.463}_{-0.320}$ & 0.196$^{+0.150}_{-0.078}$ & 0.32$^{+0.16}_{-0.10}$ & 0.98$^{+0.23}_{-0.28}$ & 0.54$^{+0.18}_{-0.16}$ \\
T   & 1961   & 1.032   & 49    & 0.41 & 8.597$^{+3.284}_{-2.634}$ & 1.394$^{+0.316}_{-0.264}$ & 0.29$^{+0.03}_{-0.06}$ & 0.99$^{+0.20}_{-0.19}$ & 0.49$^{+0.08}_{-0.10}$ \\
\hline
\noalign{\smallskip}
\end{tabular}\\
{\small Notes: All spectra were fitted with {\tt XSPEC}, using a photoelectric absorber ({\it phabs}) and a 2-temperature CIE model ({\it apec}, \citet{smith01}) and fixing the \hi\ column density $N_{\rm H}$ to an average value of $1.1\times10^{21}$\,cm$^{-2}$ \citep{church99,lebo06}. In the following, subscripts refer to the lower and higher temperature component. Col.\,(1): ID of the spectral extractions regions shown in Fig.~\ref{f2}a. All regions are point source-removed and trace the pure diffuse emission. Col.\,(2): The number of counts within the extracted region. Cols.\,(3)--(5) list the reduced $\chi^{2}$, the degree of freedom (dof), and the null hypothesis probability (nhp) of the best fit. Col.\,(6): Normalization constants of the fit. Col.\,(7): $kT$ is the mass-weighted average plasma temperature.}
\end{table*}

\paragraph{Bubble B2:}
Bubble B2 is located south of B1 at the southern end of NGC604, right above the SNR GKL98-94 (source 342 in Fig.~\ref{f3}). Its morphological appearance is very similar to bubble B1. This region also contains stars from cluster A and is clearly detected in soft and medium X-rays.
\paragraph{Bubble B3:} Bubble B3, the clearest-defined bubble, is visible in \ha\ as a luminous limb-brightened closed shell on the western side of NGC604 and is filled with soft X-ray emitting gas. In this case the observed morphology suggests that the warm ionized gas surrounds the HIM and possibly prevents it from breaking out into the ambient medium. No shell fragmentation is seen, as might be caused for example by high internal gas pressure or thermal conduction due to direct exposition of the warm gas to the HIM. It also appears unlikely that outflow has yet occurred, because the motion of the gas should follow the density gradient of the ambient medium, that is towards the northeast and away from the \ha\ wall in the west. Since there are no signs of a breakout, it is justified to consider the HIM to be bounded by the cooler \ha\ gas. The other two bubbles, especially B1, seem to have released their hot gas into cavities C1\,--\,C3. 
 
Optical data show bubble B3 to be marginally expanding, with expansion velocities of about $v_{\rm exp}\le 5$\,km/s \citep{teno00}. This suggests that the expansion of the bubble produces only a moderately strong forward shock as it sweeps up the ambient ISM. This is also supported by enhanced \sii/\ha\ and \oiii/\ha\ line ratios along the outer edge of the shell \citep[see Fig.~1 in][]{teno00}. All other line ratios for B3 are low ($<$0.33) and imply low excitation. Electron densities are likely $\le 100$\,cm$^{-3}$ \citep{apel04} and therefore within the low density limit. Fig.~\ref{f3}b also shows that B3 emits significant X-rays at medium energies from a ring-like structure. If this structure represents the limb-brightened edges of a bubble, B3 also seems to possess (similar to B1 and B2) an additional temperature component which is hotter than the one traced by the soft emission. A higher temperature would also be consistent with our above assumption that the hot gas is still trapped in B3.

\subsection{Spectral Analysis}
\begin{figure*}[t]
\includegraphics[width=16.5cm,height=10.5cm,clip,angle=0]{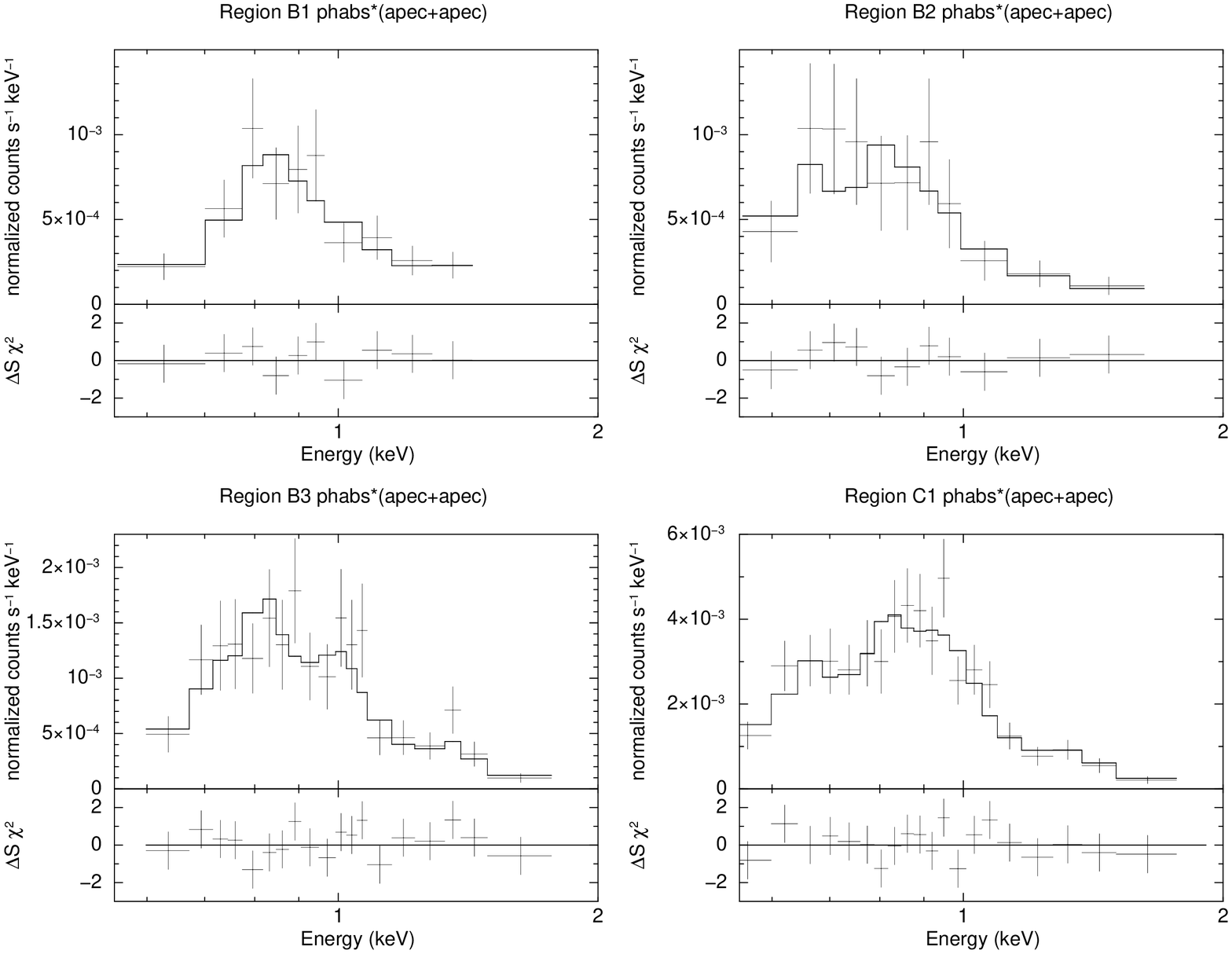}
\caption{\label{f4}
Spectra extracted from regions highlighted in Fig.~\ref{f2}a. They cover the energy range from 0.35\,-\,2.0\,keV.}
\end{figure*}

\begin{figure*}[t]
\includegraphics[width=16.5cm,height=10.5cm,clip,angle=0]{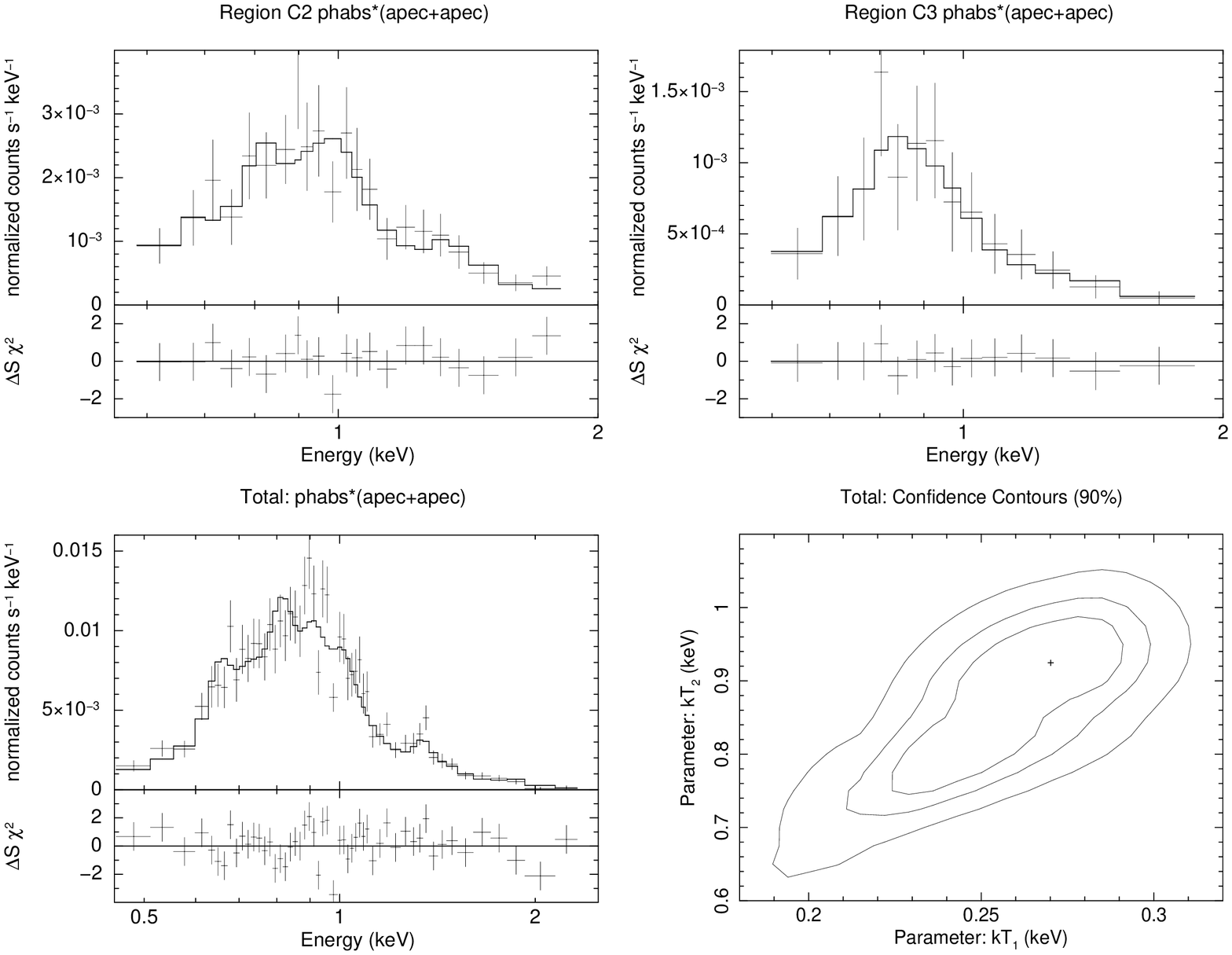}
\caption{\label{f5}
Spectra extracted from regions shown in Fig.~\ref{f2}a cover the range from 0.35\,-\,2.0\,keV, except the one extracted from the total region as it also contains emission beyond 2\,keV. $90\%$-confidence contours of ($kT_{1},kT_{2}$) are plotted for the total diffuse emission.}
\end{figure*}

Despite an exposure time of about 300\,ks, the total counts within the individual regions are relatively low (see Table~\ref{t1}). 
We therefore grouped the spectra shown in Figs.~\ref{f4} and \ref{f5} to attain a $S/N$\,$\ge$\,3.
All spectra are fitted with {\tt XSPEC} assuming a photoelectric absorber ({\tt phabs}), to correct for absorption along the line of sight, and a thermal plasma model. The underlying assumption of a thermal gas is substantiated by the fact that none of the alternative models (which consisted of a photoelectric absorber plus up to two powerlaw models) could even roughly approximate the observed spectrum (reduced $\chi^2$\,$>$\,3.8). 

For the thermal component, we started with a single CIE \citep[{\tt APEC} model, ][]{smith01}, but this resulted in statistically unacceptable fits with reduced $\chi^2$\,$>$\,2.5. The quality of the fit could be substantially improved by adding a second {\tt APEC} model (reduced $\chi^2$\,$<$\,1.1).
In addition, we experimented also with a non-equilibrium ionization (NEI) model \citep{Bork01}, as the underlying physics of a fast adiabatically expanding and continuously shocked wind might also hint at NEI conditions of the gas. Compared to the two component CIE model, the NEI-fits produced in all cases consistent temperatures and very similar $\chi^2$ values, differing by less than $10\%$. However, since the HIM is unlikely to be represented by a constant temperature, as the NEI model assumes, but rather by a temperature distribution, we considered the multi-temperature CIE model to best represent the thermal plasma.	

The \hi\ column density was set to $N_{\rm H}$ = 1.1$\times10^{21}\,{\rm cm}^{-2}$ \citep[see][]{church99,lebo06} and not allowed to vary during the fitting process. 
Relative element abundances for NGC604 were taken for He, N, O, and S from \citet{diaz87} and \citet{vil88}, and the Ne abundance value was adopted from \citet{crock06}. All other abundances were left at their revised solar values \citep{asp06} and scaled by $Z=0.65Z_{\odot}$, the value of the relative oxygen abundance. The best-fit statistics for each region are provided in Table~\ref{t1} and a representative 90\%-confidence contour plot for the ($kT_{1},kT_{2}$) parameter combination is given in Fig.~\ref{f5} for the total diffuse emission of NGC604. The average plasma temperature $kT$ was calculated by weighting the two temperature components $kT_{i}$ ($i$\,=\,$1,2$) with a factor of $\sqrt{K_{i}}/(\sqrt{K_{1}}+\sqrt{K_{2}})$, i.e., an approximate mass-weighting of the temperatures. The average plasma density $n_{\rm e}$ is simply the sum $n_{1}+ n_{2}$.

As can be seen from the total diffuse spectrum (Fig.~\ref{f5}), the diffuse X-ray gas can be reasonably well fitted with a thermal two-temperature plasma model (reduced\,$\chi^{2}$\,=\,$1.03$) which provides strong evidence that the bubbles are filled with coronal gas. 
The total, absorption-corrected, diffuse X-ray luminosity of NGC604 (0.35\,--\,2.5\,keV) is $L_{\rm X}=1.43^{+0.10}_{-0.06}\times 10^{36}$\,erg\,s$^{-1}$. However, this is only 14\% of the total diffuse X-ray luminosity of 30\,Dor \citep{town06} and emphasizes the difference between the two largest \hii-regions in the Local Group.

\section{Discussion}
\subsection{The Origin of the X-ray Emission}
There is striking evidence that NGC604 contains a very young ($\sim$3\,Myr) and rich ($\sim$200) O-type stellar population \citep[e.g.,][]{dodo81,dris93,hunt96,gonzo00,bruh03}. It is also generally accepted that NGC604 is a giant stellar wind-blown mass loss bubble \citep[e.g.,][]{rosa82,yang96,gonzo00} powered by these stars. If this is true, the young stellar population should be responsible for the X-ray gas mass and luminosity. In addition, the X-ray emission should be spatially correlated with stellar associations as the thermal plasma is considered to be produced by a shocked stellar wind \citep[e.g.][]{weaver77,shull95}. 

With this work we provide an important test of the 'stellar wind'-hypothesis by investigating which of the regions outlined in Fig.~\ref{f2}a are currently being powered by mass loss from stars. The spectral fits allow us to constrain densities and filling factors of the HIM and to calculate gas masses and luminosities for the individual regions. By constraining the mass loss history of NGC604, we estimate the stripped-off gas mass and check if it is consistent with the observed X-ray mass. Finally, we derive the luminosity expected from the standard bubble model \citep{chu95,shull95} and investigate whether it agrees with our observations. 

In cases of B1, B2, and C2 we find good correlations between regions containing massive O-stars and X-ray bright regions. Given the \ha\ and diffuse X-ray distribution, B1, B2, and C1 could be connected, implying that these bubbles were produced by the combined stellar winds from O and WR-stars sweeping up the ambient medium. In this regard, the X-ray enhancement along the \ha\ ridge in C1 could be produced by a reverse shock in the wind after impinging on the \ha\ wall. However, \citet{gonzo00} argue that the mechanical wind luminosity is sufficient to form bubbles B1 and B2 but is too low to account for the creation of C1. 
We therefore speculate that C1 is an older structure formed by stellar winds from a previous star forming event; additional heating may have been provided by SNe.

Furthermore, as blowout from regions B1, B2, and C1 is confirmed by optical spectroscopy \citep{teno00} the X-ray emission seen in cavity C3 could be due to hot gas streaming from B1 and B2 through C1 up into C3. Such motions are expected to be supersonic and should produce some characteristic emission when the X-ray gas interacts with the cooler \hi\ and \hii\ gas. The emission feature at medium energies in C3 could well represent one of these signatures. Like the X-ray enhancement near the \ha\ ridge, this harder emission could be caused by a reverse shock after the upstreaming hot plasma hits the \ha-'cap' which seems to confine NGC604 towards the north.

What kind of source powers bubble B3 and produces the X-ray emission: SNRs or stellar winds from O or WR-type stars? Perhaps the puzzling off-centered X-ray feature detected in the hard energy band at RA\,=\,01$^{h}$34$^{m}$$35^{s}$ and Dec\,=\,+30\degr47\arcmin13\farcs93 can provide some hints. This emission is significant ($S/N$\,=\,3), is well aligned with the shell, and coincides with the low excitation region observed in the optical \citep{teno00,apel04}. There are net 12\,cts in the hard band (see also Fig.~\ref{f1}) which are centered at about 3.7\,keV. Given that the Poisson probability of having 12 or more background counts in the same region is $\ll 1\%$, it is highly unlikely that this feature is caused by a background fluctuation. The reality of this feature is further substantiated by the fact that it is visible in two of our four ObsIDs. Unfortunately, the counting statistics are too low to obtain a decent spectrum and to reliably constrain its nature. 
The multi-wavelength data shown in Fig.~\ref{f3} and the 8.4\,GHz radio continuum map presented by \citet{church99} do not reveal an obvious source which can be associated with this emission. Although a contribution from SNRs cannot be ruled out, the fact that the emission is relatively hard and peaks at $\sim$\,3.7\,keV makes these sources rather unlikely candidates.

If this emission originates from NGC604 its luminosity should be in the range of $L_{\rm X}=0.9-1.2\times 10^{34}$\,erg\,s$^{-1}$ (using the Portable Interactive Multi-Mission Simulator ({\tt PIMMS v3.9d})\footnote{http://cxc.harvard.edu/toolkit/pimms.jsp} and assuming power law photon indices $\ge1.0$). This luminosity is consistent with the one expected from an X-ray binary, a pulsar wind nebula, or an AGN.
If this source is not part of NGC604 it could be an AGN (seen through the shell of B3) with an unabsorbed flux in the 2.0\,--\,8.0\,keV energy band of $F$\,$\approx$\,2.5$\times 10^{-15}$\,erg\,s$^{-1}$\,cm$^{-2}$. This flux is based on a peak line of sight \hi\ column density of $N_{\rm H}=1.5\times 10^{21}$\,cm$^{-2}$ \citep{new80} and matches that of a typical AGN detected by the \cxo\ deep field surveys. 
Considering that NGC604 covers $\sim$\,$4\times 10^{-4}$\,square degrees on the sky, 0.84 AGN are expected to cover this area \citep{Brandt05} which corroborates the AGN hypothesis. If we include the other AGN candidate $\approx$\,17$\arcsec$ eastward of B3 \citep{plu08}, the area increases to $\sim$\,$8.8\times 10^{-4}$\,square degrees and 1.8 AGN are expected in this field.  

\begin{table*}[pht]
\caption{\label{t2} Derived physical parameters based on fits presented in Table~\ref{t1}}
\begin{tabular}{cccccccccc}
\hline\hline
\noalign{\smallskip}
ID  & $T$                    & $F_{\rm X,abs\ (unabs)}$       & $L_{\rm X,abs\ (unabs)}$       & $V$           & $M_{\rm X}$  &  \multicolumn{3}{c}{$n_e\ ({\rm cm}^{-3})$} \\
\noalign{\smallskip}
    & MK                     & $10^{-14}\,{\rm erg\ s^{-1}\,cm^{-2}}$ & $10^{35}\,{\rm erg\ s^{-1}}$ & $10^{60}\,{\rm cm^{3}}$& $M_{\sun}$ & \multicolumn{3}{c}{$f_{\rm X}$} \\
(1) & (2)                    & (3)                            & (4)                            & (5)           & (6)          & 0.1  & 0.5  & 0.8\\
\noalign{\smallskip}
\hline
\noalign{\smallskip}
C1 & 4.64$^{+0.70}_{-0.58}$ & 1.96 (2.98) $^{+0.30}_{-0.40}$ & 3.90 (5.94) $^{+0.60}_{-0.80}$ & 15.5$\pm$7.75 & 2760$\pm$1040 & 0.65 & 0.29 & 0.23 \\ 
C2 & 6.38$^{+1.51}_{-0.46}$ & 1.15 (1.61) $^{+0.25}_{-0.27}$ & 2.29 (3.21) $^{+0.50}_{-0.54}$ & 5.56$\pm$2.78 & 1160$\pm$490  & 0.77 & 0.34 & 0.27 \\
C3 & 5.10$^{+2.67}_{-2.44}$ & 0.38 (0.52) $^{+0.13}_{-0.14}$ & 0.76 (1.04) $^{+0.26}_{-0.28}$ & 4.62$\pm$2.31 &  610$\pm$220  & 0.46 & 0.20 & 0.17 \\
B1 & 3.25$^{+4.52}_{-2.55}$ & 0.30 (0.42) $^{+0.21}_{-0.26}$ & 0.60 (0.84) $^{+0.42}_{-0.52}$ & 1.40$\pm$0.28 &  630$\pm$190  & 1.65 & 0.74 & 0.58 \\
B2 & 3.83$^{+1.28}_{-0.93}$ & 0.46 (0.72) $^{+0.16}_{-0.28}$ & 0.92 (1.43) $^{+0.32}_{-0.56}$ & 0.88$\pm$0.18 &  320$\pm$140  & 1.34 & 0.60 & 0.47 \\
B3 & 6.26$^{+2.09}_{-1.86}$ & 0.59 (0.82) $^{+0.18}_{-0.26}$ & 1.18 (1.63) $^{+0.36}_{-0.52}$ & 3.17$\pm$0.63 &  590$\pm$260  & 0.68 & 0.31 & 0.24 \\
T  & 5.68$^{+0.93}_{-1.16}$ & 4.86 (7.16) $^{+0.51}_{-0.28}$ & 9.68 (14.3) $^{+1.02}_{-0.56}$ & 31.1$\pm$8.58 & 5780$\pm$1670 & 0.67 & 0.30 & 0.24 \\
\noalign{\smallskip}
\hline
\noalign{\smallskip}
\end{tabular}\\
{\small Notes: Col.\,(1): ID of the spectral extraction regions shown in Fig.~\ref{f2}a. Col.\,(2): Mass-weighted average plasma temperatures. We converted from keV to K using $T\ ({\rm K})=1.16\times 10^7 kT\ ({\rm keV})$, with $kT$ being the energy of a thermal source (see Table~\ref{t1}). Cols\,(3) and (4): Absorbed and unabsorbed fluxes and luminosities, listed for the 0.35\,--2.5\,keV energy band. All uncertainties are given on a 95\%-confidence level using the {\tt error} command in {\tt XSPEC}. X-ray luminosities assume a distance to M33 of $D=817$\,kpc \citep{freed01}. Cols\,(5)\,--\,(6): Estimated volume and X-ray gas mass. $M_{\rm X}$ is calculated from electron densities if a filling factor of $f_{\rm X}=0.8$ is adopted.}
\end{table*}

The 'stellar wind'-hypothesis remains to be discussed for region B3. The multi-color image in Fig.~\ref{f3}a shows a small off-centered stellar association and several other stars close to the shell of B3, though projection effects may need to be considered. The location of the association at the SE edge of the shell (see '*' symbol in B3 of Fig.~\ref{f2}b) is also coincident with the detected FUV emission feature in the archival \galex\ data shown in Fig.~\ref{f3}c. 
The \galex\ FUV filter has its peak transmission at about 1500\AA, which traces stellar emission down to an effective temperature of $\sim$2$\times 10^4$\,K. Such a temperature roughly corresponds to stars with spectral type B0 or earlier. 
Moreover, the existence of young hot stars in B3 is also confirmed by \citet{apel04} who suggest that B3 contains several late OB stars. However, it appears questionable that B3 is the result of current stellar wind activity, simply because the number of stars seems to be too low. The current energy input of these stars makes breakout of hot X-ray gas into the surrounding medium very unlikely. Therefore, we conclude that the energy which is required to create this bubble and produce the hot gas was mainly provided by previous stellar winds or by SNe. 
We will investigate B3 and the other regions in detail in the following sections and test whether stellar mass loss is sufficient to account for the observed X-ray luminosity and gas mass. We will also determine whether the required number of stars is consistent with observations.

\subsubsection{The X-ray Gas Mass}
The determination of the X-ray gas mass allows us to investigate whether NGC604 is indeed powered by $\sim 200$ O/WR-type stars whose heavy mass loss drives a steady wind into the ambient ISM. For that purpose, we assume that the derived X-ray luminosities listed in Table~\ref{t2} are entirely produced by a reverse shock hitting the wind and heating the gas up to X-ray temperatures. This is the emission mechanism assumed by the standard model of stellar wind-blown bubbles \citep{cas75,weaver77,shull95}.

The determination of the X-ray gas mass also requires knowledge of the emitting volume $V$, the X-ray filling factor $f_{\rm X}$, and the electron density $n_{\rm e}$. These variables are related to the normalization constant $K$ of the spectral fit as follows:
\begin{equation}
K=\frac{10^{-14}}{4\pi D_{\rm A}^{2}}\int n_{\rm e} n_{\rm H}\,dV,
\end{equation} 
where $D_{\rm A}$ is the angular size distance to the source in (cm), and $n_{\rm e}$ and $n_{\rm H}$ are the electron and hydrogen densities per cm$^{-3}$, respectively. The ionized gas mass traced by X-rays can be written as:
\begin{equation}
M_{\rm X}= 1.15\ m_{\rm H} n_{\rm e}\,V f_{\rm X}, 
\end{equation}
where $1.15\times m_{\rm H}$ is the mass per hydrogen atom, accounting for the contribution of He and assuming a fully ionized gas with solar abundances \citep{asp06}.

As these quantities cannot directly be measured, reasonable assumptions need to be made. We are aware that these assumptions can be a serious source of uncertainty, especially in the case of $V$ where most of the uncertainty is due to poorly constrained viewing angles, non-spherical geometries, projection effects, and (most importantly) unknown linear dimensions along the line of sight. However, despite these uncertainties the following calculations provide valuable insights into the energetics and the evolution of NGC604. Moreover, most of the above mentioned uncertainties can be minimized, e.g., by a visual inspection of the available multi-wavelength data and by providing reasonable uncertainty ranges. 

In the following we assume that bubbles B1\,--\,B3, the clearest defined structures in NGC604, can be approximated by spheres (see Fig.~\ref{f2}a) with radii\footnote{The dimensions were determined from the \hst\ {\it WFPC2} \ha\ image, which provides a spatial resolution of $\sim$\,0\farcs1 (=\,0.36\,pc).} of $r_{\rm B1}=22.5$\,pc, $r_{\rm B2}=19.3$\,pc, and $r_{\rm B3}=29.6$\,pc, respectively. If the bubbles are spherically symmetric, the uncertainties associated with the radius would be negligible ($<$\,4\%). However, to account for possible asymmetries and for uncertainties in the distance to \m, we assume an uncertainty in the volume of 20\%. 

Estimating the volumes of the cavities C1\,--\,C3 is much harder because of their seemingly irregular shapes. However, if we consider that NGC604 is embedded into a large reservoir of cold gas \citep[e.g.,][]{apel04,tosa07} through which star formation propagates and carves out the blister-like cavities, one can approximate such regions by more regular geometries. After an inspection of the \hst\ \ha\ image we decided to approximate cavity C1 by a box of about 140\,pc\,$\times$\,64\,pc\,$\times$\,64\,pc in size. In order to correct for possible projection effects and the unknown 'real' extent along the line of sight, we adopt a 50\% uncertainty in the volume (which will also be the default value for C2 and C3). Because region B1 seems to be part of C1, we subtracted the volume of B1 from the volume of C1. 
Cavity C2 can be approximated by a box which measures $\sim$47\,pc\,$\times$\,86\,pc\,$\times$\,47\,pc, whereas C3 is represented by a box whose extent amounts to about 47\,pc\,$\times$\,58\,pc\,$\times$\,58\,pc. 
All resulting volumes are listed together with their uncertainties in Table~\ref{t2}.

We now proceed to constrain the filling factor of the hot phase and derive the electron density within the individual regions. Unfortunately, both quantities are not known a priori, but as follows from Eq.\,(2), $n_{\rm e} \sim f_{\rm X}^{-1/2}$, which allows us to calculate the density for different filling factors. A high value of $f_{\rm X}$ is implied for the bubbles, because the X-ray plasma seems to be confined by the cooler gas (cf. Fig.~\ref{f3}a). Furthermore, a filling factor of 0.1 was derived by \citet{gonzo00} for the warm ionized gas in NGC604. This leaves $\sim$10\% for the warm neutral and molecular medium; a reasonable value if one takes into account that the stellar wind and radiation field ionized and swept away most of the neutral gas within these bubbles. This argument is confirmed by the IRAC 8$\mu$m image shown in Fig.~\ref{f3}d, in which all bubbles appear to be relatively free of dust. 
A high value of the X-ray filling factor $f_{\rm X}$ is also expected for the cavities, because the shocked wind-driven gas should be able to easily fill a large volume (in the present evolutionary state cooling is negligible, see below). Therefore, it appears reasonable to assume a high filling factor of $f_{\rm X}=0.8$ for NGC604 (densities for $f_{\rm X}=1.0$ are up to 0.04\,cm$^{-3}$ lower). 

The resulting electron densities and gas masses are given in Table~\ref{t2}. Mass-related uncertainties are calculated by propagating the errors for volume, filling factor, and density, taking into account that $\Delta n_{i}\sim~\sqrt{\Delta K_{i}/(\Delta f_{\rm X} \Delta V)}$. The uncertainty associated with the filling factor was set to $\Delta f_{\rm X}$\,=$\,\pm0.1$. 
To calculate $M_{\rm X,T}$, the total X-ray emitting mass in NGC604, we simply used the sum of the individual volumes. The mass integrated over the individual regions is consistent with the total mass within the calculated uncertainty ($\sum M_{\rm X,ID} = 6100\pm1200\,M_{\odot}$ vs. $M_{\rm X,T}=5800\pm1700\,M_{\odot}$).

\subsubsection{The Stellar Mass Loss History}
In this section we investigate whether the observed stellar population can account for the X-ray emission by its mass loss history and which regions are likely to be powered by these sources. This assumes that the X-ray emission is produced by a reverse shock hitting the stellar wind. Answering this question requires detailed knowledge about the number and type of resident stars, their age, and their mass loss history.  

Photometric observations of NGC604 that were conducted with the \hst\ revealed that about 186 stars are located in bubbles B1 and B2. The majority of these stars are of type O9.5 or younger \citep[cluster A in][]{hunt96,bruh03}. These stars are considered to have an age of 3\,Myr and formed during an instantaneous starburst \citep{gonzo00}. Among these 186 stars, about 14 are classified as spectral type WR/Of \citep{dris93,bruh03}. For simplicity, we assume constant mass loss and neglect evolutionary effects as the stars move off the main sequence. 

A typical mass loss rate of an O9.5 star with an initial mass of 40$\,M_{\odot}$ is $\dot{M}=1.1\times 10^{-6}M_{\odot}$/yr \citep{lam93} and of a WN7 star with the same initial mass is $\dot{M}=3.8\times 10^{-5}M_{\odot}$/yr \citep{nug00}. We also made the assumption that the average time spent in the WR-phase amounts to $\sim$2.58$\times 10^5$\,yr \citep{maed94}.
With these values, we obtain $\sim$$750\,M_{\odot}$ of material that has been released by the stars. This value needs to be corrected by the significant amount of gas which flows into the cavity from the shell boundaries due to evaporation via thermal conduction. In order to estimate this effect we make the simplifying assumptions of a constant plane-parallel flow, neglecting magnetic suppression ($\kappa_{0}$\,=\,1) and turbulent mixing, and assume that the evaporated gas is immediately heated up to X-ray temperatures. 

Following Shull \& Saken (1995), the mass gain rate due to thermal evaporation can be written as:
\begin{equation}
\dot{M}_{\rm e} \approx 2 \frac{\mu}{k} R_{\rm b}\,C\,T_{\rm b}^{5/2} \kappa_{0}, 
\end{equation}
where $\mu$\,=\,$0.62\,m_{\rm H}$ is the mass per particle assuming a fully ionized plasma with 10\% helium and $K(T)$\,=\,$C\,T_{\rm b}^{5/2}$ is the thermal conductivity with $C=6\times 10^{-7}$\,erg\,s$^{-1}$\,cm$^{-1}$\,K$^{-7/2}$. {The temperature of the bubble can be approximated as:
\begin{equation}
T_{\rm b} = 5.3\times 10^6\ L_{38}^{8/35}\,n_{0}^{2/35}\,t^{-6/35}\kappa_{0}^{-2/7}\ \mbox{K}, 
\end{equation}	
and $R_{\rm b}$, the radius of the bubble, is given by:
\begin{equation}	
R_{\rm b}= 65.9\ L_{38}^{1/5}\,n_{0}^{-1/5}\,t_{6}^{3/5}\ \mbox{pc}.
\end{equation}	
Here $t_{6}$ represents the time in units of 1\,Myr and $n_{0}$ is the ambient gas density per cm$^3$. The mechanical luminosity of the wind (in units of $10^{38}$\,erg\,s$^{-1}$) can be expressed by: 
\begin{equation}
L_{\rm 38}= 0.0126 N_*\,\dot{M}_6\,v_{2000}^2, 
\end{equation} 
where $N_*$ is the number of O-type stars having initial masses $\ge$\,$40M_{\odot}$, $\dot{M}_{6}$ is the mass loss rate in units of $10^{-6}M_{\odot}$/yr and $v_{2000}$ represents the terminal wind velocity in units of 2000\,km/s.
Consequently, the mechanical luminosity produced by the O9 and WR-stars in B1 and B2 amounts to $L_{\rm 38,O9}=2.38$ \citep[$v_{2000}=1.0$,][]{lam93} and $L_{\rm 38,WR}=3.10$ \citep[$v_{2000}=0.68$,][]{nug00}, respectively, which totals $L_{\rm 38}=5.48$.

Setting $n_{0}=1.0$\,cm$^{-3}$ \citep{rosa84}, $\kappa_{0}=1$, and $t_{6}=3$ yields $R_{\rm b}=179$\,pc and $T_{\rm b}=6.5\times 10^6$\,K. These values are inconsistent with the observed ones and imply an evaporated mass of about $M_{\rm e}=25,000\,M_{\odot}$ in 3\,Myr, a value which is more than 20 times larger than the estimated X-ray gas mass in B1 and B2. \citet{apel04} also found a mismatch between observationally derived quantities and predictions from the standard bubble model. They argued that the external pressure from the ambient ISM and outflows of hot gas could slow down the expansion of the bubbles. 

In order to resolve the observed inconsistencies we adopt for B1 and B2 a mass-weighted temperature of $T_{\rm b}=3.45\times 10^6$\,K and a bubble radius of $R_{\rm b}=26.5$\,pc (based on $V_{\rm B1+B2}=2.28\times10^{60}$\,cm$^{-3}$ and a spherical geometry). Under these assumptions, the evaporated gas mass amounts to $\sim$770\,$M_{\odot}$ which, together with stellar mass loss, yields $\sim$1500\,$M_{\odot}$ in 3\,Myr. This is still much higher than the total observed gas mass of $\sim$$950\pm250$\,$M_{\odot}$. 

Because we rule out significant contributions from SNe (see Sect. 4.2) stellar mass loss remains the favored mechanism to account for the X-ray gas.} One way to reconcile these results is to consider the possibility that some of the X-ray emission in cavities C1 and C3 is from gas originating from the stellar wind of clusters A and B which streamed upwards and now fills the remaining cavities.  In order to investigate this, we now have to include the additional 8 O9-type stars from cluster\,B \citep{hunt96} in our calculations. Among the 194 stars, about 15 are classified as spectral type WR/Of \citep{dris93,bruh03}. The 179 O-stars alone lose approximately $590\,M_{\odot}$ of their mass within 3\,Myr, whereas the 15 WR-stars, most of which appear to be type WN7/WN8 \citep{dodo81}, release $150\,M_{\odot}$ into the ambient medium. Because WR-stars evolve from O-stars, we need to estimate the mass loss from the progenitor star, too. Within 2.74\,Myr this amounts to about $45\,M_{\odot}$. The 194 O/WR stars should have lost about 800\,$M_{\odot}$ during their lifetime which together with an evaporated gas mass of $M_{\rm e}=3200 M_\odot$ totals $\sim$$4000 M_{\odot}$.
This mass is in good agreement (within 7\%) with the integrated mass of $4300\pm1100$\,M$_{\odot}$ in regions B1, B2, C1, and C3. It should be noted that the evaporated gas mass is a conservative upper limit, because the Spitzer conductivity does not take into account saturation effects and magnetic fields which can significantly suppress conduction.

A third important mass loading effect, photoevaporation from molecular clouds, can be neglected. Although a large molecular gas reservoir has been reported for NGC604 \citep[e.g.,][]{blitz85,apel04}, radiative cooling in bubbles with radii $r\ge 30$\,pc is expected to be unimportant in view of the low density of the hot gas \citep{shull95}.

We think it is possible in principle to get matching X-ray gas masses if cavity C2 is also included. However, there are no indications that B1 and B2 are connected to C2. It is conceivable that the X-ray emission seen in C2 is in part produced by a stellar wind emanating from B2 caused by some stars of cluster A and in part by three smaller star clusters (see Fig.~\ref{f2}b for their positions). Moreover, the faint FUV emission seen in C2 (in Fig.~\ref{f3}c) is also indicative of stellar continuum and supports the existence of young massive stars. 
In the western part of NGC604 only $\sim$20\% of the gas mass stems from mass loss. If we assume an even lower fraction of 5\% for C2, 60$M_{\odot}$ need to be produced by stellar mass loss, which would require 18 O-type stars (assuming $t_6=3$ and $\dot{M}_{6}=1.1$). However, these stars are not observed in the HST images of this region, and so the gas is unlikely to be produced by the current generation of stars there.  It is possible that the gas is left over gas from previous stellar winds and that SNe re-energized the gas and provide additional heating over time. High resolution spectrophotometric data are needed to investigate the stellar populations and its age in the eastern part of NGC604 to estimate their contribution to the overall X-ray emission.

As also noted by \citet{apel04}, the case of bubble B3 is different from B1 or B2 because this region might be shielded by the \ha\ ridge from the stellar wind of the central cluster. Interestingly, this bubble is expanding at very slow rate, indicating that this structure is much older than, e.g., regions westwards of the \ha\ ridge \citep{teno00,apel04}. Adopting again the same age and mass loss rate as above and assuming that 5\% ($\sim$30$M_{\odot}$) of the X-ray emission from B3 are produced by a shocked stellar wind, 9 O9-type stars would be needed to account for the observed X-ray gas mass. Considering that this bubble contains only a small unresolved stellar association and $\sim$3 isolated stars, presumably O-stars or red supergiants, the number of stars can hardly account for the X-ray gas mass. 
We therefore speculate that the hot gas in bubble B3 stems from a previous stellar wind and/or was heated by SNe. Because SNRs expanding into a low density gas are hard to detect \citep{mk87,chu90}, the lack of detected SNRs in this region is not surprising. As a result of the expected low energy input and marginal expansion velocity of B3, we also expect the bubble to remain stable and confine the hot plasma in its interior.

It should be noted that there is a significant spread in the reported value of $\dot{M}$ for the O/WR-stars, depending on the adopted model. For our calculations we assumed a simple homogeneous wind \citep{lam93,maed94} rather than a clumped one \citep{nug00,full06}. However, even in the case of a clumped wind in which the mass loss rate for O stars can be one order of magnitude lower, consistent gas masses are obtained within the uncertainties.

\subsubsection{X-ray Luminosities}
If the hot gas in NGC604 is produced by a shocked stellar wind, the observed X-ray luminosities should also be in agreement with theoretical predictions from corresponding models \citep[e.g.,][]{shull95,chu95}. We therefore calculate the expected X-ray luminosities for each region starting with the main volume which consists of B1, B2, C1, and C3. We assume again that the observed X-ray emission is the result of the shocked stellar wind.

The X-ray luminosity of a bubble can be written as:
\begin{equation}
L_{\rm X}=(8.2\times 10^{27}\mbox{erg\,s}^{-1})\,n_{0}^{10/7} R_{\rm b}^{17/7} v_{\rm e}^{16/7} I(\tau)\,\xi, 
\end{equation} 
where $v_{\rm e}$\ 
is the expansion velocity of the bubble, $\xi$ is the metal abundance in solar units. and $\tau$ is the ratio of the minimum temperature and the central temperature of the bubble.
$I(\tau)$ is a dimensionless integral which can be approximated by: $I(\tau)=3.79-5\tau^{1/2}+(5/3)\tau^3-(5/11)\tau^{5.5}$, with $\tau=0.16\,L_{37}^{-8/35} n_{0}^{-2/35} t_{6}^{6/35}$ 
\citep[see][for details]{chu95}. 

With $L_{37}=58$ it follows that $\tau=0.084$ and hence $I(\tau)=2.34$. These values together with a spectroscopically determined expansion velocity of $v_{\rm e}\approx36$\,km\,s$^{-1}$ \citep[slits (b)\,--\,(e) in][]{rosa84,yang96} yield a predicted X-ray luminosity of $L_{\rm X}=8.70\times 10^{35}$\,erg\,s$^{-1}$. This value is well within the uncertainty limit of the unabsorbed X-ray luminosity integrated over the western part of $L_{\rm X}=9.25^{+0.84}_{-1.14}\times 10^{35}$\,erg\,s$^{-1}$. If a clumped stellar wind is assumed the corresponding luminosity amounts to $L_{\rm X}=8.42\times 10^{35}$\,erg\,s$^{-1}$.
This finding together with a matching gas mass fully supports the wind-blown bubble scenario for the western part of NGC604 and does not require additional heating from SNe.

In a similar way we derived luminosities for the slowly expanding regions B3 and C2 \citep[$v_{\rm e}\approx10$\,km\,s$^{-1}$,][]{teno00}. They amount to $L_{\rm X, B3}=6.3\times 10^{33}$\,erg\,s$^{-1}$ and $L_{\rm X,C2}=1.1\times 10^{34}$\,erg\,s$^{-1}$, respectively. These values underestimate the observed luminosities by about a factor of 30 and clearly rule out current stellar mass loss as the main contributor of the X-ray emission in these regions. It also corroborates our previous claim that the stellar populations in C2 and B3 are not powerful enough to account for the determined gas mass. Although mass loss of the current stellar generation could account for approximately 4\% of the observed X-ray luminosity in C2 and B3, it appears more likely that most of the mass of the hot gas was produced by previous stellar wind activity with possible additional heating by SNe.

It is important to note that we also checked how the X-ray luminosities and gas masses obtained for the western and eastern part of NGC604 change if the $N_{\rm H}$ is fixed to a lower (Galactic) value of 5.21$\times10^{20}\,{\rm cm}^{-2}$ \citep{dl90} and to an upper value of 1.3$\times10^{21}\,{\rm cm}^{-2}$ \citep{new80}. In both cases the results were consistent with those obtained for the fixed average $N_{\rm H}$ of 1.1$\times10^{21}\,{\rm cm}^{-2}$. The maximum deviations are 14\%. In addition we also allowed the $N_{\rm H}$ to vary during the fits. The fitted values of the $N_{\rm H}$ were poorly constrained with large uncertainties and a large spread in the derived values. For four regions (C2, B2, B3, and T) the fitted values of the $N_{\rm H}$ were consistent with 1.1$\times10^{21}\,{\rm cm}^{-2}$. For regions C1 and C3, the fitted values (0.24$\times10^{21}\,{\rm cm}^{-2}$ and 0.11$\times10^{21}\,{\rm cm}^{-2}$) were less than the Galactic column and therefore considered unreliable. For region B1, the fitted value was 3.83$\times10^{21}\,{\rm cm}^{-2}$, which is larger than the total column through \m\ and therefore also unreliable. Given the poor constraints on $N_{\rm H}$ from the X-ray spectral fits, we decided to adopt the results of the spectral fits which had assumed a fixed value of 1.1$\times10^{21}\,{\rm cm}^{-2}$. However, the luminosities and gas masses for an unconstrained $N_{\rm H}$ were always within the range of uncertainty provided by the fits which assumed a lower and upper $N_{\rm H}$ of 5.21$\times10^{20}\,{\rm cm}^{-2}$ and 1.3$\times10^{21}\,{\rm cm}^{-2}$, respectively.

\subsection{Contributions from SNe}
We need to constrain the number of SNe to evaluate whether they contribute significantly to the energy injected into the ISM and to the determined X-ray gas mass. This kind of investigation can only be done reliably if high resolution spectrophotometric data for all regions are available. Unfortunately, the available data restrict a detailed exploration of the stellar population exclusively to the western part of NGC604. \citet{bruh03} investigated about 50 OB-stars in cluster A covering the mass range from $20\le M_{\odot}\le 150$. They assumed a Salpeter IMF with $\alpha=-2.3$ and compared the predicted number of stars with their observations. In the mass range between $80\le M_{\odot}\le 120$ two stars would have been expected, but none were detected. We therefore assume that these stars , if formed, exploded as SNe during the last 3\,Myr. Interestingly, the 3 most massive ($120\le M_{\odot}\le 150$) and extremely rare stars still seem to exist which would imply that they formed during the last 3\,Myr. However, these authors do not rule out that these stars are unresolved massive binaries. 

Because the above mentioned study covers only $\sim$25\% of the massive stellar population in cluster A, we need to scale the expected number of SNe accordingly. Since the number of exploded stars of the remaining population is unknown and may also vary from region to region, we simply extrapolate linearly and obtain a conservative upper limit of 8 SNe during the last 3\,Myr. 

The total thermal energy of the gas in the western part provided by stellar wind activity amounts to about $2.5\times10^{52}$\,erg, which is roughly equivalent to the energy input of about 25 SNe. This number of SNe is three times higher than our estimate of the expected number of SNe and is therefore inconsistent with the assumption that SNe contributed significantly to the heating of the gas. Besides the fact that the X-ray luminosity can be well explained by stellar mass loss alone, there is another aspect which argues against contributions from SNe. The starburst models from \citet{leit92} show that the total energy input from stars formed during an instantaneous starburst 3\,Myr ago, is entirely provided by stellar winds. This lets us assume that no or only a few SNe occurred during the lifetime of the current stellar generation and that mass load by SNe is negligible. Although there could have been a couple of SNe, they would not have provided significant heating.

The thermal energy of the X-ray gas from stellar winds amounts to $\sim$$5.0\times10^{51}$\,erg and is five times smaller than the total thermal energy. This discrepancy could be explained, e.g., by radiative cooling due to excessive mass injection by evaporation \citep{shull95} or by energy losses from the hot interior due to outflowing gas.
Because the stellar population in the eastern part is completely unconstrained, we refrain from making predictions of the expected number of SNe.

\subsection{The overall picture}
In X-rays as well as at optical wavelengths a clear dichotomy is seen in NGC604. The western part seems to be a young ($\sim$\,3\,Myr) and dynamically active region in which ongoing stellar mass loss is most likely responsible for the observed X-ray gas mass and luminosity. This part of NGC604 is X-ray bright compared to the bubbles of the LMC studied by \citet{chu90} and \citet{wang91}. Such X-ray bright bubbles usually require additional energy input from SNRs in order to reconcile the observed luminosity with the predicted one from the standard bubble model. Hence, the western hemisphere of NGC604 seems to be a new case in which an X-ray-bright bubble is produced by stellar mass loss and the current stellar population either has not formed SNe, or any SNe so far were inefficient in heating the gas.	

The eastern hemisphere appears to be an aged structure in which the gas seems to be in a dynamically relaxed condition. This relaxation could be caused by the lack of ionizing sources and the presence of the \ha\ ridge which most likely shields this region from the east. 
The likely reason why this part of NGC604 is also X-ray bright and has by far the highest temperatures is that radiative cooling is negligible in regions of this size \citep{shull95,teno00}. It also implies that the gas is indeed confined to C2 and B3.
Contrary to the west, the observed X-ray luminosity in C2 and B3 cannot be explained by current stellar mass loss alone. In this case, off-center SNRs could produce the additional emission \citep{chu90}. 

Due to the lack of identified young clusters, the X-ray emission in C3 most likely does not originate from local stellar clusters; clusters C and D \citep{hunt96} are too far off-center to contribute significantly. Therefore, the emission in C3 could originate from the shocked stellar wind expelled by clusters A and B. If this is correct it also follows that the ISM is highly porous, because the overpressured hot gas can stream freely some 200\,pc from B1 through C1 into C3 and fill large cavities. Such supersonic motions can explain the large filling factors of the HIM and are a characteristic feature in models of wind-blown bubbles.
 
These streaming motions should also cause characteristic shock features when the supersonic wind interacts with the gas of the shell wall. 
The medium X-ray emission along the \ha\ ridge and at the \ha-'cap' seen in the contour map (Fig.~\ref{f2}) and in Fig.~\ref{f3} can be interpreted as fingerprints of such shocks. This shock scenario is consistent with the crossing timescale of the stellar wind which emerges from clusters A and B and traverses cavity C1, which is of the order of $\sim$$5\times 10^5$\,yr, assuming that the wind slows down from originally $v_{2000}=1$ to a conservative value of $v_{2000}=0.1$. A little less than 1\,Myr is required for the wind to reach the top of cavity C3 (assuming the same wind speed). These timescales are much shorter than the cluster age, and therefore, if SNe and SNRs can be ignored, the wind could get continuously shocked over a period of several million years.

\section{Summary and conclusions}
We presented the first X-ray images of NGC604, that are deep enough to reveal the complex interplay among different constituents of the ISM. A strong negative morphological correlation was found between \ha\ and X-ray emitting gas. All bubbles and cavities are filled with hot coronal gas ($3.3-6.4\times 10^6$\,K), suggesting high volume filling factors of about $f_{\rm X}=0.8$ and densities $n_{\rm e}<0.6$\,cm$^{-3}$.

A total, absorption-corrected, diffuse X-ray luminosity in the soft band (0.35\,--\,2.5\,keV) of $L_{\rm X}=1.43\times 10^{36}$\,erg\,s$^{-1}$ was found for NGC604, which is about 14\% of the corresponding value for 30\,Dor \citep{town06}. The X-ray gas mass in NGC604 amounts to about 6000\,$M_{\odot}$, which is only 15\% of the corresponding mass in 30\,Dor \citep{wang99}, and emphasizes the differences between the two largest \hii-regions of the Local Group even more.

The sum of the detected X-ray gas mass in the western part of NGC604, containing regions B1, B2, C1, and C3, amounts to $\sim$$4300\pm1100$\,M$_{\odot}$ and the unabsorbed X-ray luminosity of $L_{\rm X}=9.25^{+0.84}_{-1.14}\times 10^{35}$\,erg\,s$^{-1}$ places this region into the regime of X-ray bright bubbles. Both, the X-ray gas mass and the luminosity agree well with expectations from the standard bubble model and are consistent with mass loss from about 200 O/WR-stars. In view of an estimated cluster age of 3\,Myr, the consistency between observed and expected luminosities, and the predictions from the starburst models from \citet{leit92}, the current stellar generation seems to be too young for SNe to have made a significant contribution to heating. 
The western hemisphere contradicts the general view in which the luminosity of X-ray bright bubbles is underestimated by the standard model and additional contributions from SNRs are required. The X-ray gas likely originates from the stellar O-type population in clusters A and B. We find that mass load from evaporation due to thermal conduction contributes about 95\% to the X-ray gas mass.

In the eastern hemisphere the estimated mass of X-ray emitting gas amounts to $\sim$$1750\pm550$\,M$_{\odot}$. The sum of the predicted X-ray luminosity from regions B3 and C2 which one would expect from a shocked stellar wind during the lifetime of the donor stars, largely underestimates the observed unabsorbed luminosity of $L_{\rm X}=4.84^{+0.62}_{-0.75}\times 10^{35}$\,erg\,s$^{-1}$ by about a factor of 30. This result supports the hypothesis that the luminosity of X-ray bright bubbles is inconsistent with the one predicted by the standard bubble model and argues for additional contributions from SNRs. We rule out a shocked stellar wind produced by the current stellar generation as the main source of the X-ray emission. We cannot rule out that the gas was created during an era when the stellar population was powerful enough to drive a continuous wind and/or SNe heated the gas. The number of expected SNe in C2 and B3, respectively, remains unconstrained due to the lack of spectrophotometric data of the stellar population.

The dichotomy between east and west was also found for the warm ionized gas. \citet{teno00} noted that the dynamics and ionization of the gas in the eastern part of NGC604 is completely different from the western part. As in the case of the HIM, the \ha\ ridge appears to play an important role by separating both hemispheres. The eastern region seems to be an old part of NGC604 in which the warm gas appears to be in a dynamically relaxed condition. In contrast, the western region is dominated by young stars which create a highly dynamical ISM via strong stellar winds and contribute essentially $100\%$ to the X-ray gas mass and luminosity.  
The freely-streaming winds in NGC604 should produce characteristic emission features when they interact with the shell walls. Therefore, we interpret the medium X-ray emission along the \ha\ ridge and at the \ha-'cap' as direct fingerprints of such shocks. Provided this is correct, these spots of enhanced emission would confirm the expected supersonic motions of the gas. 

\acknowledgments
This work has made use of SAOImage DS9 \citep{joy03}, developed by the Smithsonian Astrophysical Observatory, the FUNTOOLS utilities, and the HEASARC FTOOLS package. We thank the anonymous referee for her/his valuable comments. RT wishes to thank J. Raymond and H. Lee for helpful discussions and suggestions. RT acknowledges support under NASA \cxo\ award number GO6-7073A. TJG and PPP acknowledge support under NASA contract NAS8-03060. PG acknowledges support through \cxo\ grant GO6-7073B and PFW acknowledges support through G06-7073C.

\end{document}